\documentclass[fleqn,twoside]{article}
\usepackage{espcrc2}
\usepackage{amssymb}

\usepackage{graphicx}
\usepackage{epsfig}
\newcommand{\AmS}{{\protect\the\textfont2
  A\kern-.1667em\lower.5ex\hbox{M}\kern-.125emS}}

\newcommand{\be}{\begin{equation}}
\newcommand{\ee}{\end{equation}}
\newcommand{\beqn}{\begin{eqnarray}}
\newcommand{\eeqn}{\end{eqnarray}}
\newcommand{\eq}[1]{(\ref{#1})}

\title{
Static $\bar{Q}$-$Q$ Potential from $N_f$$=$$2$ Dynamical Domain-Wall QCD
}

\author{Koichi Hashimoto\address{Institute for Theoretical Physics, 
Kanazawa University, Kanazawa 920-1192, Japan}
\hspace*{-1.5mm}${}^{,}$\hspace*{-1mm}
\address{Radiation Laboratory, RIKEN, %(The Institute of Physical and Chemical Research), 
Wako 351-0198, Japan}
and Taku Izubuchi${}^{\mathrm{a,}}$\address{RIKEN-BNL Research Center, 
Brookhaven National Laboratory, Upton, New York 11973, USA}
for RBC Collaboration
\thanks{
This work is supported by RIKEN Super Combined Cluster at RIKEN. 
We thank RIKEN, Brookhaven National Laboratory and the U.S. Department of 
Energy for providing the facilities essential for the completion of this work. 
K.H. thanks RIKEN BNL Research Center for 
its hospitality where this work was performed.}
}
\begin{document}

\begin{abstract}
We calculate the static quark and anti-quark potential both in 
quenched and two-flavor dynamical quark lattice QCD using DBW2 gauge and domain-wall 
quark actions. Lattice spacings from Sommer scale are determined.
We find (i) mixing of excited states is different in between quenched and dynamical, 
(ii) lattice spacing $a_{r_0}\sim a_{m_{\rho}}$ in dynamical and
(iii) coefficient of Coulomb term being $\alpha_{N_f=0}<\alpha_{N_f=2}$ at 
$a^{-1}\sim$ 2 GeV.
\end{abstract}

% typeset front matter (including abstract)
\maketitle

\section{Introduction}
This study has the following purposes: 
(i) determination of lattice spacing $a$ from Sommer scale $r_0=R_0a\approx$0.5 
fm~\cite{Sommer scale}, 
(ii) observation of dynamical quark effects in the static potential, 
such as 
larger coefficient of Coulomb term 
and string breaking.
All quoted errors are statistic only, and systematical errors will be reported 
elsewhere \cite{RBC}.

\section{Calculation}
The static potential $V(\vec{R})$ between infinitely heavy quark 
and anti-quark separated by $\vec{R}$ in spatial direction 
is obtained from the Wilson loop $\langle W(\vec{R},T) \rangle$: 
\begin{equation}
\langle W(\vec{R},T) \rangle = C(\vec{R}) e^{-V(\vec{R})T} + {\rm (``excited~states")},
\label{eq:Wilson loop}
\end{equation}
where $C(\vec{R})$ is the overlap with the ``ground state" 
which is normalized with $C(\vec{R}=\vec{0})=1$.

We calculate $\langle W(\vec{R},T) \rangle$ both 
in quenched and dynamical QCD. We employ quenched DBW2 gauge action~\cite{DBW2 glue} 
with $\beta$=0.87, 1.04 on $16^3\times32$ and 
$\beta$=1.22 on $24^3\times48$ lattices (100, 405, 106 configurations respectively), 
corresponding to $a^{-1}_{m_{\rho}}\sim$1.3, 2 and 3 GeV~\cite{DBW2,DBW2 3GeV}, 
and DBW2 and $N_f=2$ domain-wall fermion~\cite{DWF} dynamical action with 
$\beta=0.80$, $L_s$=12, $M_5$=1.8, $m_{\rm dyn}a$=0.02, 0.03, 0.04 
(941, 559, 473 configurations respectively), corresponding to 
$m_{\pi}/m_{\rho}$=0.53(1), 0.60(1), 0.65(1) on $16^3\times32$ lattices.
Inverse of the lattice spacing from rho meson mass in the chiral limit is 
$a^{-1}_{m_{\rho}}\sim$1.7 GeV in dynamical~\cite{Dynamical DWF}.

We implement Bresenham algorithm~\cite{Bresenham algorithm}, 
which allows us to obtain the geodesic path connecting the ends of 
$\vec{R}=(N_1,N_2,N_3)$ on lattice, 
where $N_i$ ($i=1, 2, 3$) are any three integers, 
and APE smearing~\cite{APE smearing} for spatial links.
The smearing coefficient and iteration time are tuned: $(c,n)=(0.50, 20\sim25)$ for both 
dynamical and quenched to maximize $C(\vec{R})$.

The static potential $V(\vec{R})$ and the overlap with the ground state $C(\vec{R})$ are obtained 
from~\eq{eq:Wilson loop}: 
\begin{eqnarray}
V(\vec{R},T)&=&\ln\left[\langle W(\vec{R},T) \rangle/\langle W(\vec{R},T+1) \rangle\right], \\
C(\vec{R},T)&=&\langle W(\vec{R},T) \rangle^{T+1}/\langle W(\vec{R},T+1) \rangle^T
\end{eqnarray}
by neglecting the excited states.
$T$ is selected as the smallest time 
on which the excited states contribution becomes negligible in $V(\vec{R})$.

The physical parameters are obtained from a fitting function:
\begin{eqnarray}
V(\vec{R})&=&V_0-\frac{\alpha}{R}+\sigma R,~~R=|\vec{R}|, \\
R_0&=&\sqrt{\frac{1.65-\alpha}{\sigma}}.
\end{eqnarray}

\begin{figure}[h]
\begin{center}
\includegraphics[angle=-00,scale=0.40,clip=true]{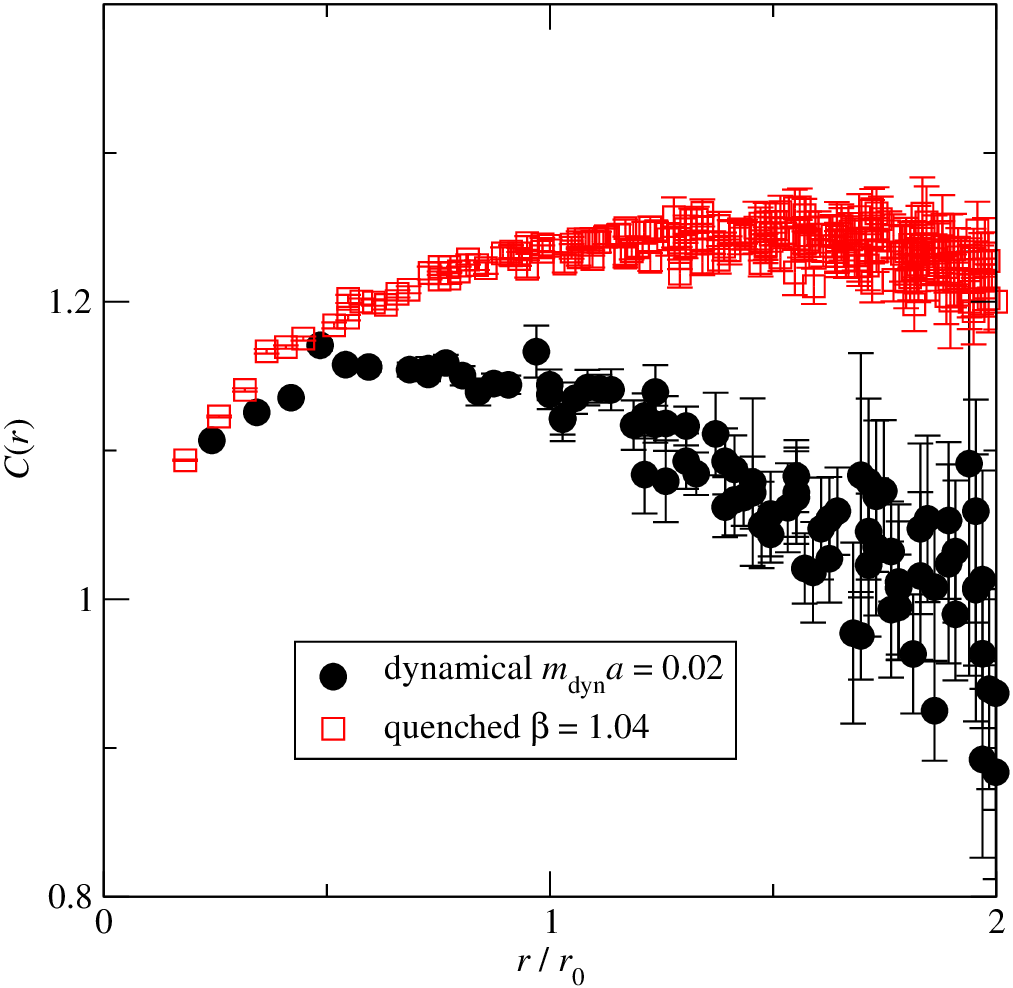}
\end{center}
\vspace{-14mm}
\caption{$C(r)$ vs. $r/r_0$ for dynamical ($m_{\rm dyn}a=0.02$) 
and quenched ($\beta=1.04$) extracted at $T$=5.}
\label{fig:overlap}
\vspace{-8mm}
\end{figure}

\section{Results}

Figure~\ref{fig:overlap} shows the overlap with the ground state $C(\vec{r})$, 
where $\vec{r}$=$\vec{R}a$, in dynamical and quenched at $T=5$.
In quenched $r$ dependence is small at $r \geq r_0$.
On the other hand $C(\vec{r})$ decreases as $r$ increases in dynamical.
This suggests the mixing of excited states is different for dynamical 
and quenched. 
Further investigation about the excited states in dynamical case is needed.
Note that $C(\vec{r})>1$ may mean the existence of excited states with 
negative norm which is possible
because improved gauge actions break reflection positivity~\cite{Luecher Weisz} 
at small $T$.

\begin{figure}[t]
\begin{center}
\includegraphics[angle=-00,scale=0.37,clip=true]{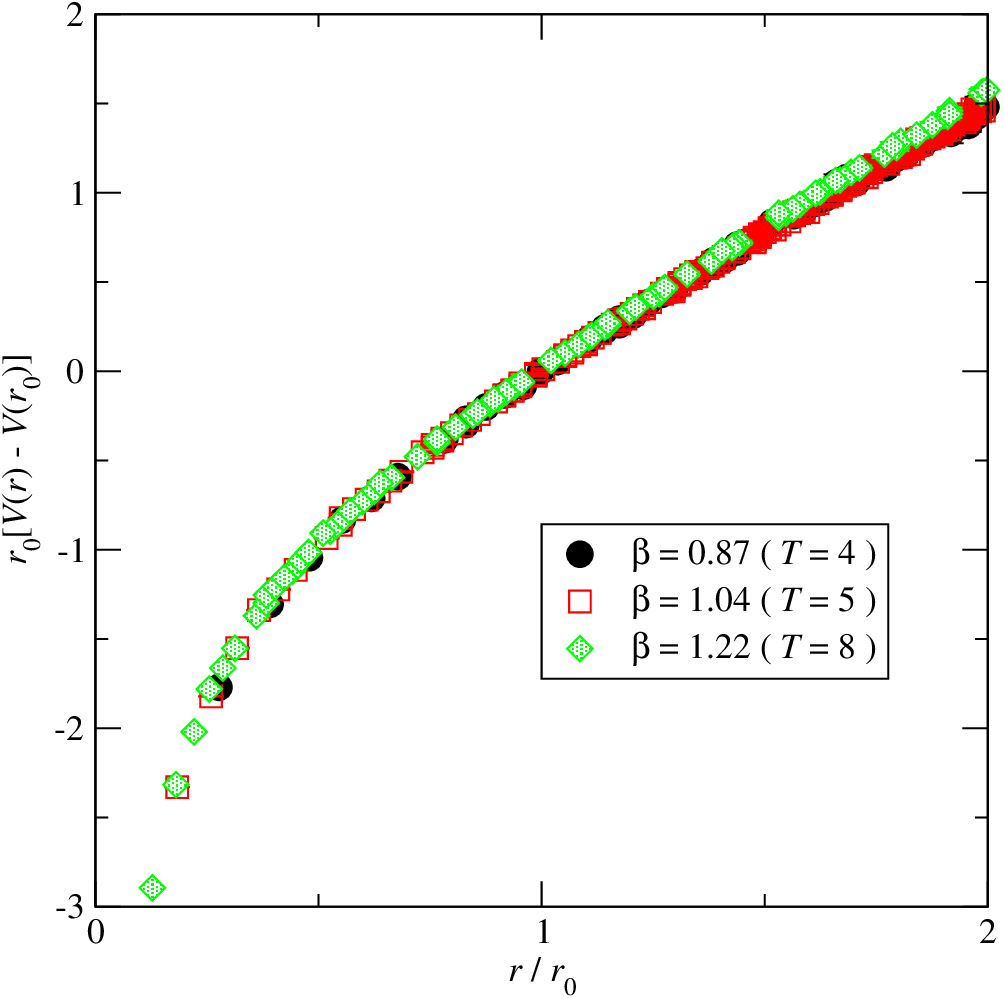}
\includegraphics[angle=-0,scale=0.37,clip=true]{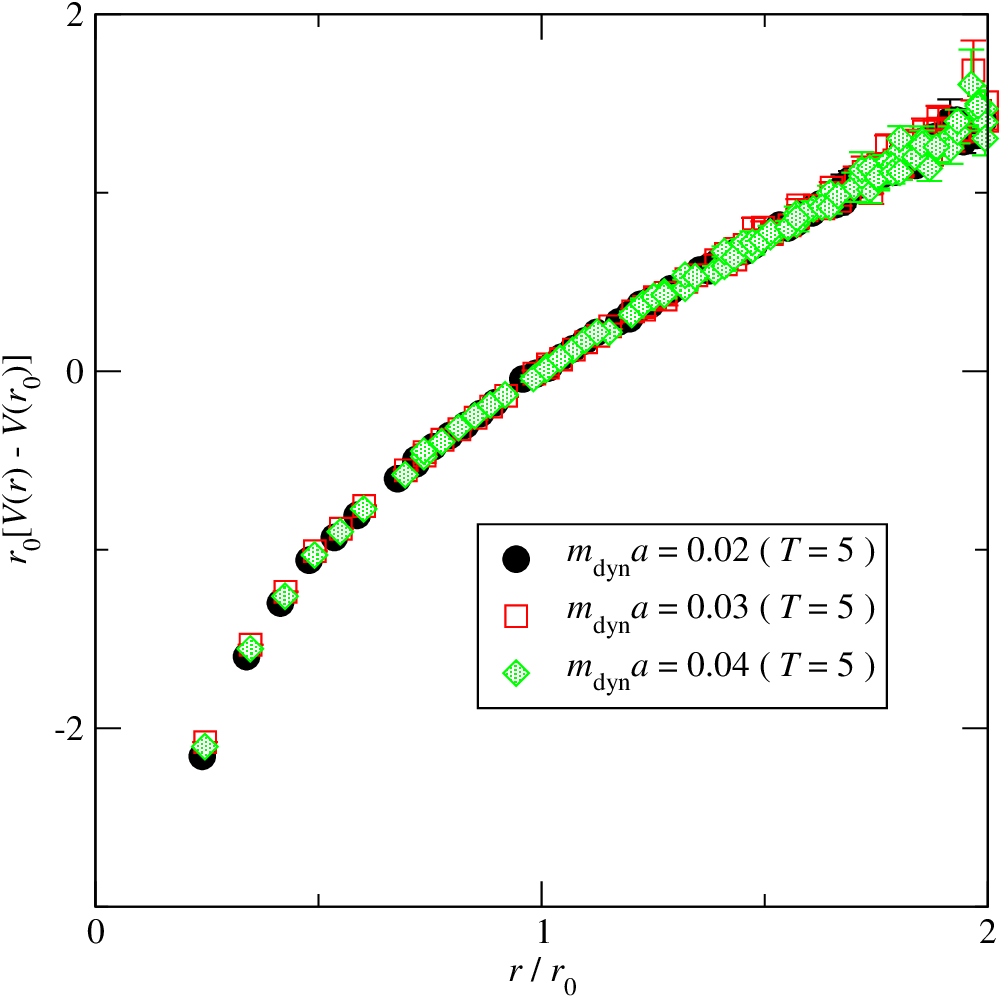}
\end{center}
\vspace{-14mm}
\caption{$r_0[V(r)-V(r_0)]$ vs. $r/r_0$ for quenched (upper) 
and dynamical (lower).}
\label{fig:pot}
\vspace{-6mm}
\end{figure}

Figure~\ref{fig:pot} shows the static potential for quenched (upper) and 
dynamical (lower), normalized with Sommer scale $r_0$: 
$r_0[V(r)-V(r_0)]$ vs. $r/r_0$. 
$T$ is selected 4, 5, 8 for $\beta=0.87$, 1.04, 1.22 respectively 
in quenched, and $T=5$ in dynamical. 
When scale is fixed by $r=r_0$, the scaling violation 
is very small in quenched. 
We did not see any sign of the string breaking in dynamical: 
the potential at $r\gg r_0$ increases linearly in current statics.

\begin{figure}[h]
\begin{center}
\includegraphics[angle=-00,scale=0.40,clip=true]{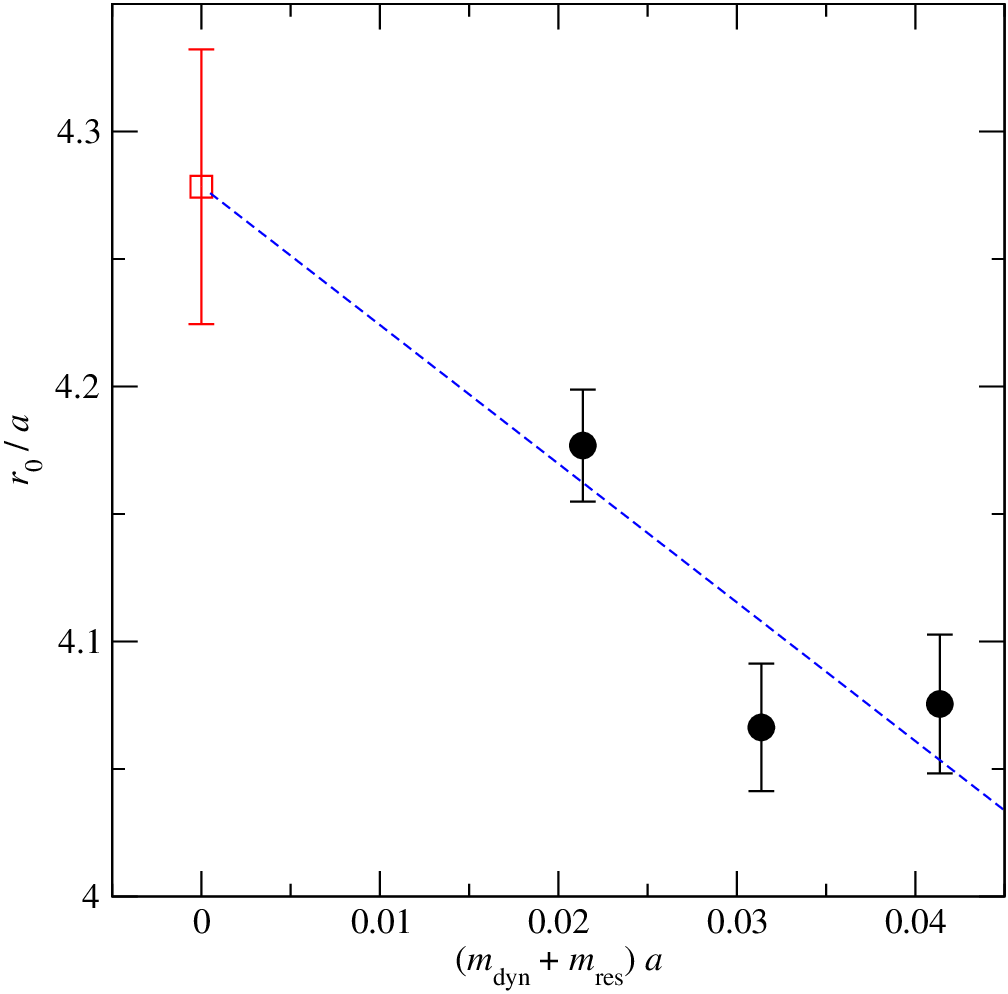}
\end{center}
\vspace{-10mm}
\caption{$r_0/a$ vs. $(m_{\rm dyn}+m_{\rm res})a$ 
with linearly extrapolate the chiral limit.}
\label{fig:r0_dyn}
\vspace{-4mm}
\end{figure}

Figure~\ref{fig:r0_dyn} shows the Sommer scale $R_0$=$r_0/a$ for different dynamical 
quark mass, $m_{\rm dyn}$+$m_{\rm res}$, 
where $m_{\rm res}a$=0.00137(4)~\cite{Dynamical DWF}.
We obtain 
$R_0=4.28(5)$ 
in the chiral limit,
$m_{\rm dyn}+m_{\rm res}\to 0$, 
by linear extrapolation. Therefore lattice spacing is 
$a^{-1}_{r_0}$=1.69(2) GeV with Sommer scale $r_0$=0.5 fm.

\begin{table}
    \begin{center}
        \begin{tabular}{|c|c|c|} \hline
             & $a^{-1}_{r_0}$ [GeV] & $a^{-1}_{m_{\rho}}$ [GeV]\\ \hline
            dyn. in the chiral limit  & 1.69(2) &1.69(5)~\cite{Dynamical DWF} \\ \hline
            quenched $\beta=0.87$ & 1.43(1) & 1.31(4)~\cite{DBW2} \\ \hline
            quenched $\beta=1.04$ & 2.15(1) & 1.98(3)~\cite{DBW2} \\ \hline
            quenched $\beta=1.22$ & 3.09(2) & 2.91(5)~\cite{DBW2 3GeV} \\ \hline
        \end{tabular}
        \caption{$a^{-1}_{r_0}$ vs. $a^{-1}_{m_{\rho}}$. Quenched $\beta=1.22$ is preliminary. 
        $\beta=0.87$ and 1.04 are reanalyzed in this work after \cite{DBW2}}.
        \label{tbl:spacing}
    \end{center}
    \vspace{-10mm}
\end{table}

We compare lattice spacing from Sommer scale and that from rho meson mass 
(Table~\ref{tbl:spacing} ).
$a^{-1}_{r_0}$ is a few percent larger than $a^{-1}_{m_{\rho}}$ for quenched.
On the other hand $a^{-1}_{r_0}$ is consistent with $a^{-1}_{m_{\rho}}$ within the error 
in dynamical.
The lattice spacings from $r_0$ and $m_{\rho}$ become much closer by dynamical 
quark effect.

\begin{figure}[h]
\begin{center}
\includegraphics[angle=-00,scale=0.40,clip=true]{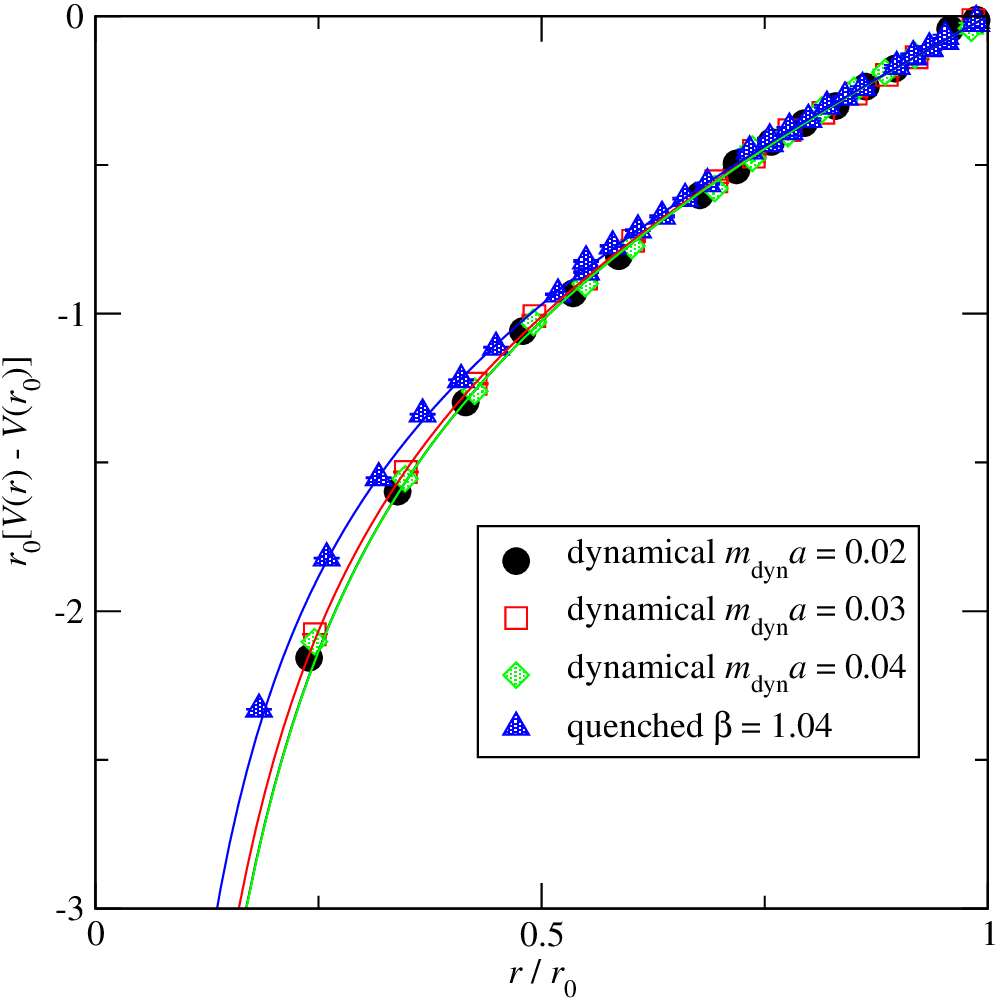}
\end{center}
\vspace{-10mm}
\caption{$r_0[V(r)-V(r_0)]$ vs. $r/r_0$ for $r/r_0\ll1$. Lines are fitted potential.}
\label{fig:pot_mini}
\vspace*{-4mm}
\end{figure}

In Figure~\ref{fig:pot_mini}, we compare the static potential in between 
quenched and dynamical at short range, 
normalized with $r_0$.
When the scale is fixed by $r=r_0$, we see a deeper Coulomb potential for dynamical 
than that of quenched. 
Also the coefficient of Coulomb term $\alpha$ increases, 
i.e. $\alpha_{N_f=0}<\alpha_{N_f=2}$.
This relation is consistent with the perturbative screening effect 
of dynamical quarks.

\section{Conclusion}

We calculated the static potential both in quenched and two-flavor dynamical domain-wall
QCD and obtained lattice spacings estimates (Table~\ref{tbl:spacing}).
We saw the following dynamical quark effects: 
(i) $a^{-1}_{r_0}>a^{-1}_{m_{\rho}}$ for quenched and 
$a^{-1}_{r_0}\sim a^{-1}_{m_{\rho}}$ for dynamical in the chiral limit, 
(ii) deeper Coulomb potential in dynamical than in quenched 
due to screening effect: $\alpha_{N_f=0}<\alpha_{N_f=2}$.

String breaking was not observed.
However we observed the different behaviors of $C(\vec{R})$ 
between quenched and dynamical.
This suggests the existence of the heavy-light excited states.

To study string breaking, we may have to treat the excited states more carefully.
Multi-state calculation~\cite{Bresenham algorithm,M-state} 
is one of the possibilities.

\end{document}